\documentclass{ifacconf}
\usepackage{amssymb}
\usepackage{amsmath}
\usepackage{graphicx}
\usepackage{amssymb}
\usepackage{bbm}
\usepackage{easylist}
\usepackage{fp}
\usepackage{siunitx}
\usepackage{tikz}
\usepackage{xcolor}
\usepackage{color}
\usepackage{multirow}
\usepackage{subfigure}
\usepackage{graphicx}      
\usepackage{natbib}

\usepackage{enumitem}

\newtheorem{df}{Definition}
\newtheorem{RM}{Remark}

\begin{document}
\begin{frontmatter}


\title{Truck Platoon Formation at Hubs: \\ An Optimal Release Time Rule}

\thanks[footnoteinfo]{This work is supported by the Strategic Vehicle Research and Innovation Programme through the project Sweden4Platooning, Horizon2020 through the project Ensemble, the Swedish Foundation for Strategic Research and the Swedish Research Council. The work of Ehsan Nekouei is supported by the start-up grant 7200658 from City University of Hong Kong.}


\author[First]{Alexander Johansson} 
\author[First]{Valerio Turri} 
\author[Second]{Ehsan  Nekouei} 
\author[First]{Karl H. Johansson} 
\author[First]{Jonas M\aa rtensson}

\address[First]{KTH Royal Institute of Technology, Stockholm, Sweden (e-mail: \\ \{ alexjoha, turri, kallej, jonas1 \}@ kth.se).}
\address[Second]{City University of Hong Kong, 
   Kowloon, Hong Kong (e-mail: enekouei@cityu.edu.hk)}

\begin{abstract}   
We consider a hub-based platoon coordination problem in which vehicles arrive at a hub according to an independent and identically distributed stochastic arrival process. The vehicles wait at the hub, and a platoon coordinator, at each time-step, decides whether to release the vehicles from the hub in the form of a platoon or wait for more vehicles to arrive. The platoon release time problem is modeled as a stopping rule problem wherein the objective is to maximize the average platooning benefit of the vehicles located at the hub and there is a cost of having vehicles waiting at the hub. We show that the stopping rule problem is monotone and the optimal platoon release time policy will therefore be in the form of a one time-step look-ahead rule. The performance of the optimal release rule is numerically compared with (i) a periodic release time rule and (ii) a non-causal release time rule where the coordinator knows all the future realizations of the arrival process.  Our numerical results show that the optimal release time rule achieves a close performance to that of the non-causal rule and outperforms the periodic rule, especially when the arrival rate is low.

\end{abstract}


\end{frontmatter}

\section{Introduction}

A platoon is a formation of connected vehicles driving on the road with small inter-vehicular distances. A high degree of automation ensures safety and fuel efficiency of vehicles in a platoon. The lead vehicle in a platoon is typically maneuvered by a human driver, while the follower vehicles automatically follow their respective in-front driving vehicles. 

Platooning is expected to be an important element of the future intelligent transport systems thanks to the following main benefits: First, the workload of drivers of the follower vehicles in the platoon decreases due to automation. The monetary savings of platooning can be substantial if the drivers can utilize their time for other tasks, \emph{e.g.,} administrative duties, rest, etc.  Second, the energy consumption of vehicles in platoons decreases thanks to reduced air drag. The reduced energy consumption has been validated by field experiments in \cite{Alam2015}, \cite{Browand2004} and \cite{Tsugawa2016}, where potential energy savings of around  $10\%$ were reported. Third, the traffic capacity of roads increases when vehicles drive in a synchronized manner and by the small inter-vehicular distances. Forth, the safety increases due to communication and automation in platoons.

\begin{figure}
	\centering
	\includegraphics[scale=0.28]{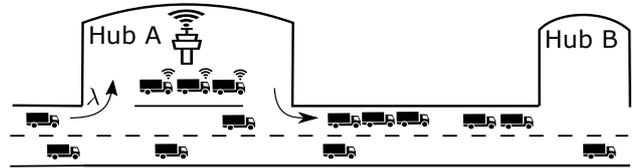}
	\caption{The platoon coordinator decides whether to release the vehicles at the hub as a platoon, or wait for more vehicles to arrive.} 	 
	\label{PMS}
\end{figure}

Platoon coordination is needed in order for vehicles to form platoons. In general, platoon coordination includes deciding which vehicles will form platoons and how each platoon will be formed. In this work, we consider the truck platooning scenario illustrated in Figure \ref{PMS}, where some vehicles have to wait at a location along a highway, called hub, in order for other vehicles to catch up so that platoons can be formed at the hub. The platoon coordination is then to decide when to release vehicles from the hub in form of platoons. In today's transportation infrastructure, there are many examples of locations which could function as hubs, \emph{e.g.,} freight terminals, gas stations, parking areas, tolling stations, harbors, etc.




Coordination for vehicle platooning has been studied in the literature. \cite{Boysen2018},  \cite{Larsen2019}, \cite{Farokhi2013},
and \cite{Johansson2018} study variations of hub-based platoon coordination problems. In the aforementioned works it is assumed that the incoming vehicles' arrival times to the hub are known to the coordinator. This assumption is valid if transport companies are willing to share the trip information and locations of their vehicles. 

\cite{Zhang2017} study platoon coordination of two vehicles under stochastic traveling times. The vehicles have a common meeting point in a three-way network. The authors showed that it is motivated to form the two-vehicle platoon if the vehicles' scheduled arrival times at the meeting point differ less than a certain threshold. Different from \cite{Zhang2017}, in this paper we consider (more than two) vehicles that arrive to a hub according to a stochastic arrival process.

\cite{Adler2016} study platoon coordination at a hub when vehicles arrive according to a stochastic arrival process. The authors showed the optimality of a threshold policy for releasing vehicles from the hub under the assumption that the arrival process is Poisson distributed. Different from \cite{Adler2016}, in this paper we show the  optimality of a threshold policy for releasing vehicles at a hub without assuming a Poisson arrival process.

\cite{Liang2016}, \cite{Larsson2015}, \cite{Hoef2018}, \cite{Xiong2020} study the platoon coordination problem where platoons are formed on the road without stopping at hubs. The main drawback of such approach is that during the formation phase, some vehicles speed up or slow down in order to merge. This may lead to traffic flow reduction and speed limit violation.  For a review of platoon coordination strategies, we refer the reader to \cite{Bhoopalam2018}.

In this work, we consider a hub-based platoon coordination problem, where a coordinator decides, at each time-step, if vehicles will be released from the hub as a platoon or they stay at the hub. The coordinator does not know the realization of the arrival process a priori. Instead, the coordinator knows the statistical distribution of the vehicle arrival process. We assume that arrivals are independent and identically distributed. The contributions of this paper are threefold:
\begin{enumerate}[label=(\roman*)]
	\item We formulate the platoon release time problem as an optimal stopping rule problem.
	\item We derive the optimal release rule, under the assumption that arrivals are independent and identically distributed. We show that the optimal rule is in the form of the one time-step look-ahead policy. The optimality follows by showing that the platoon release time problem is monotone. 
    \item We simulate a hub located along with a highway in Sweden, where the historical traffic data is used to set realistic values of the vehicle arrival rates to the hub. 
 \end{enumerate}

This paper is structured as follows. In Section \ref{SM}, we present the system model,
including the vehicle arrival process, the decision variables, the reward function and the optimization problem of the coordinator. The solutions of the optimal release time problem are presented in Section~\ref{OSP}. In Section~\ref{SIM}, the optimal release rule is evaluated in a simulation of a hub in Sweden. Finally, conclusions are provided in Section \ref{CON}.

\section{System model}\label{SM}

In this section, we formulate the hub-based platoon coordination problem that is illustrated in Figure \ref{PMS}. First, the vehicle arrival process is defined. Then, we define the coordinator's decision variables and  reward function. Last,  the optimal release time problem is formulated.

\subsection{Arrival process to the hub}

The number of vehicles that arrive to the hub at time-step $k>0$ is denoted by the random variable $X_k$. The realization of $X_k$ is denoted by $x_k\in \mathbb{Z}_{\geq0}$, which takes non-negative integer-values. At the initial time-step $k=0$, there are $n_0\in \mathbb{Z}_{>0}$ vehicles located at the hub. We assume $n_0>0$, since if there are zero vehicles located at the hub, there is no decision to make at $k=0$. The number of vehicles located at the hub at time-step $k>0$ is denoted by the random variable $N_k$ and its realization is denoted by $n_k$, which is defined by $n_k=n_{k-1}+x_k$.  The realizations $x_1,...,x_k$ are known at time-step $k$. Hence, $n_k$ is known to the coordinator at time-step $k$. Moreover, we assume the random variables $X_1,X_2,...$ to be independent and identically distributed (i.i.d) and their distribution is known to the platoon coordinator. The probability mass function of the arrivals is denoted by $\text{P}(x)=\Pr(X_{k}=x)$.

\begin{RM}
In real traffic situation, the arrival rate of vehicles to the hub is expected to be higher at peak periods than during off-peak periods. This is observed in Figure \ref{Lambdaplot}. However, the i.i.d. assumption of $X_1,X_2,...$ is justified if the traffic conditions change slowly in comparison to the time that vehicles stay at hubs. Then, different distributions can be used for the peak and off-peak periods, and the distributions can be estimated by historical data.   
\end{RM}

\subsection{Decision variables}	
 The coordinator's decision variable, at time-step $k$, is denoted by $u_k\in\{0,1\}$. If $u_k=1$, the vehicles at the hub are released as a platoon and the coordinator receives a reward (that is introduced later), and $u_k=0$ corresponds to not releasing the vehicles at time-step $k$. Exactly one release time is allowed and when the decision to release is taken, the problem terminates (and possibly starts over). The coordinator has to release the vehicles at latest at time-step $J$, which can be selected arbitrary large, \emph{i.e.,} $u_0+...+u_J=1$. The release time is not known a priori and the decision to release at time-step $k$ is a causal function of all the information up to time $k$.  Therefore, $u_k$ is a realization of a random variable, which we denote by $U_k$. 
 
 The feasible set of decisions, which the coordinator can take at time-step $k$, depends on if the problem already has been terminated. Let $h_k=u_0+...+u_{k-1}$, where $h_k=1$ if the problem has been terminated before time-step $k$ and $h_k=0$ otherwise. The variable $h_k$ is a realization of a random variable denoted by $H_k$. Given $H_k=h_k$, the feasible set of decisions at time-step $k$, is defined as
\begin{equation*}
\mathcal U_k(h_k)=\begin{cases}
\{0,1\}, &\text{ if } h_k=0 \text{ and } k<J,  \\
\{0\}, &\text{ if } h_k=1, \\ 
\{1\}, &\text{ if } h_k=0 \text{ and } k=J.  \\
\end{cases}
\end{equation*}

\subsection{Reward function}	

If a group of vehicles are released from the hub at the same time-step, they will form a platoon, as shown in Figure~\ref{PMS}. In order to define the reward function of the platoon coordinator, we make two assumptions on vehicles' benefit from platooning. First, the lead vehicle in each platoon has zero benefit. Second, the follower vehicles have equal benefits $R>0$. Then, the average platooning benefit in a platoon of $n$ vehicles is $R({n-1})/{n}$. The reward of the platoon coordinator if it releases the vehicles at time-step $k$ ($u_k=1$) is
\begin{equation*}\label{comut}
y_{k}(n_k)=\frac{n_k-1}{n_k}R-c k,
\end{equation*}
where the first term is the average platooning benefit of the $n_k$  vehicles that are released, and the second term  is the cost associated with vehicles waiting at the hub. 

\begin{RM}
	Note that in general, we might consider the reward function of the form $y_k(n_k)=f(n_k)-\Lambda(k)$, where $f(n_k)$ represents the platooning benefit and $\Lambda(k)$ represents the cost of waiting at the hub. The results presented later will hold as long as the reward function fulfills the monotonic property that will be introduced later.  
\end{RM}

\subsection{The optimal release time problem }	

The platoon coordinator aims to maximize its expected reward with respect to its decision variables $U_0,...,U_J$. The optimization problem of the coordinator is
\begin{equation}\label{opt}
\max\limits_{U_0,...,U_J}\text{E}\left[   \sum_{k=0}^{J} y_k(N_k) U_k  \right],
\end{equation}
where the event $U_k=1$ corresponds to releasing the vehicles at time-step $k$. In the next section, the solution of the optimization problem is presented.

\begin{RM}
The optimal release time problem \eqref{opt} is an optimal stopping rule problem, where releasing the vehicles from the hub corresponds to stopping. In the next section, two approaches for solving the optimal release time problem are discussed. 
\end{RM}

\section{Solutions to the optimal platoon release time problem}	\label{OSP}

 In this section, we give two solutions of the optimal release time problem. The first solution is to calculate the optimal decision in the form of a state feedback policy by means of stochastic dynamic programming. This solution is not limited to the i.i.d assumption of the vehicle arrival process which was made in the previous section. The second solution is a release time rule that looks one time-step ahead. The optimality of the release time rule is shown under the i.i.d assumption of the arrival process.

\subsection{Stochastic dynamic programming} \label{PF}

The optimal release time problem \eqref{opt} can be solved by stochastic dynamic programming. Let the random variable $S_{k}~=~(N_k,H_k)$ denote the state of the system at time-step $k$ and its realization is denoted by $s_{k}~=~(n_k,h_k)$. Given the state $s_k$, the maximal expected reward received from time-step $k$ to $J$ is denoted by $V_k(s_k)$ and is defined by  
\begin{align*}
&V_k(s_k)= \\  &  \hspace{0.6cm}\max \limits_{u_k \in \mathcal U_k(h_k)}\{ y_k(n_k)u_k+  \text{E}\left[V(S_{k+1}) |S_k=s_k,U_k=u_k \right]\},
\end{align*}
where the maximal expected reward is $V_k(s_k)=y_k(n_k)$ if the optimal decision is $u_k~=~1$. Note, $V_k(s_k)=0$ if we have $h_k=1$, since then $\mathcal U_l(h_l)=\{0\}$ for $l\geq k$. An optimal state feedback policy $u_k=u_k^*(s_k)$ for $k=0,...,J$ is calculated by setting $V_{J+1}(s_{J+1})=0$ and calculating $V_J(s_J),...,V_0(s_0)$ backwards (from $J$ to $0$) as functions of the states $s_0,...,s_J$, respectively. However, the stochastic dynamic programming becomes intractable when $J$ is large and the possible realizations of $X_1,...,X_J$ are many.

\subsection{Optimal one-step look-ahead policy}

The main result of this section is an optimal release rule where the platoon coordinator looks one time-step ahead, instead of (initially) looking $J$ time-steps ahead, as in the stochastic dynamic programming solution.  Before stating  the optimal release time rule, we define the notions of the one time-step look-ahead release rule and monotone release time problems.

\begin{df}\label{sla}
	The one time-step look-ahead release rule calls for releasing at time-step $k$ if
	\begin{equation}\label{sla1}
	y_{k}(n_k) \geq \text{E}\left[y_{k+1}\left (N_{k+1})\right | N_k=n_k \right],
	\end{equation}	
	that is, the reward of releasing vehicles at time-step $k$ is greater than the expected reward of releasing at time-step $k+1$.
\end{df}

\begin{df}[Monotone release time problem]\label{mono} 
\hspace{0cm}
The release time problem is called monotone if the event that inequality \eqref{sla1} is satisfied at time-step $k$ implies that it will also be satisfied at time-step $k+1$, for all possible realizations of $X_{k+1}$.  
\end{df}

\begin{thm}\label{thm1}
Consider the platoon release time problem \eqref{opt}. Then, the following statements hold: 
\begin{itemize}
	\item[(i)] The release time problem is monotone in sense of Definition \ref{mono}. 
\item[(ii)] The one time-step look-ahead release rule  is optimal and releasing is optimal at the first time-step $k$ such that $n_k\geq n^*$, where 
	\begin{equation*} \label{correq}
	n^*=\text{min}\{n\geq0|\frac{c}{R} \geq \sum \limits_{x} \frac{x}{n^2+nx}\text{P}(x) \},
	\end{equation*}
	where $\text{P}(x)$ denotes the probability of $x$ vehicles arriving to the hub at one time-step.
	\end{itemize}
\end{thm}

\begin{pf}
See appendix \ref{AA}.	
\end{pf}

\begin{RM}
The threshold $n^*$ is a function of the cost-benefit ratio $c/R$ and the distribution of random arrivals $X_1,...,X_J$.
\end{RM}

	  \begin{RM}
The optimal state feedback policy of the coordinator is  $u_k~=~u_k^*(s_k)$, where 
\begin{equation*}
u^*(s_k)  =  \begin{cases} 1 \text{ if } n_k \geq n^* \text{ or } k=J, \text{ and if }  h_k  =  0 \\
0 \text{ else,} \end{cases} 
\end{equation*}
where the criteria $h_k=0$ imposes that releasing the vehicles at time-step $k$ requires that the coordinator has not released the vehicles before time-step $k$.
\end{RM}
	  

\section{Numerical results}\label{SIM}

In this section, we simulate the optimal one time-step look-ahead release rule to study the average utility of vehicles, the average platoon length and the average waiting times under different arrival rates for a potential hub in Sweden. We start by describing the simulation setup.

\subsection{Setup of simulation}

\begin{figure}	
	\centering
	\includegraphics[scale=0.612]{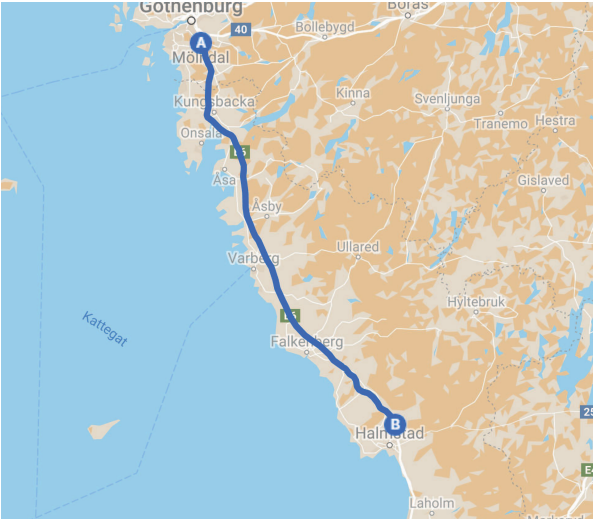}
	\caption{Map of a region in Sweden. The location of the hub is denoted by the letter $A$ (near Gothenburg) and the vehicles at the hub can platoon on the highway to the point that is denoted by $B$ (near Halmstad). The map is copied from Google maps. }
	\label{Karta}
\end{figure}

We consider a hub located near the city of Gothenburg, in Sweden. In Figure \ref{Karta}, the position of the hub is denoted by $A$ and the vehicles at hub $A$ can platoon to the point denoted by $B$.  The distance between hub $A$ and point $B$ is approximately $120$ km.

\begin{figure}
	\centering
	\includegraphics[scale=0.63]{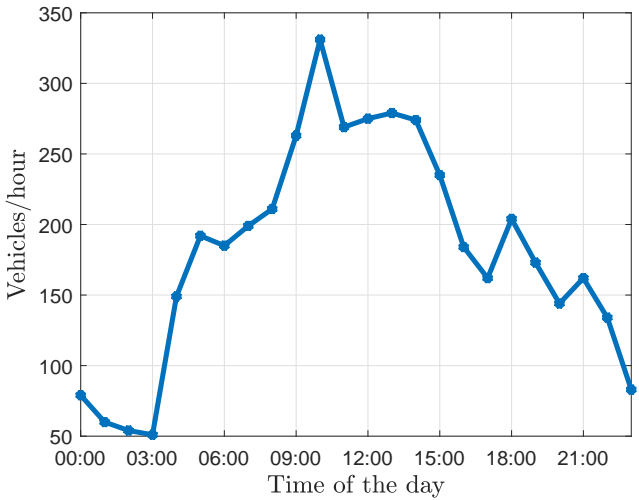}
	\caption{Average number of trucks that pass by the hub (point $A$ in Figure \ref{Karta}) during each hour of the day.}
	\label{Lambdaplot}
\end{figure}

In the simulations, we set the time-step length to $5$ seconds. At each time-step $k$, the number of arriving vehicles $X_k \sim \text{Po} (\lambda)$ is Poisson distributed with mean $\lambda$. The initial number of vehicles at the hub, $n_0>0$, is drawn according to the zero-truncated Poisson distribution.

Figure \ref{Lambdaplot} shows the average number of vehicles that passed by the hub (point $A$ in Figure \ref{Karta}) during each hour of the day over a period of ten weekdays. The data was collected by the Swedish Transport Administration. The data in Figure \ref{Lambdaplot} is used to compute realistic values of the arrival rate  $\lambda$. For example, in average 330 vehicles pass by the hub during the peak period 10:00--11:00~a.m. If $120$ of those vehicles stop at the hub  in order to platoon and if vehicles arrive to the hub according to a Poisson process then the expected number of arrivals at each time-step (per $5$ seconds) is $\lambda=1/6$. In the simulations, $\lambda$ is varied between $0$ and $1/6$.

\begin{figure}
	\centering
	\includegraphics[scale=0.64]{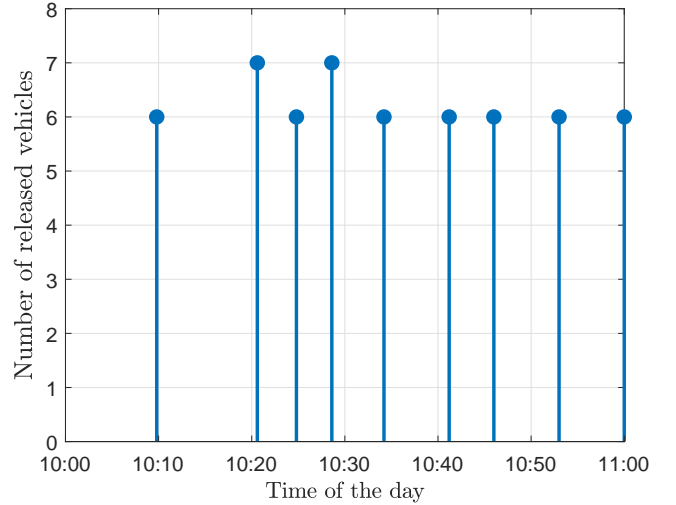}
	\caption{The number of released vehicles from the hub $A$ in the period 10:00--11:00 a.m. under the optimal release rule and the arrival rate $\lambda=1/6$.}
	\label{Illtime}
\end{figure}

 Figure \ref{Illtime} shows the number of vehicles released from hub $A$ by the coordinator under the optimal release rule for a realization of the arrival process between 10:00--11:00 a.m. In this figure, the arrival rate to the hub is $\lambda=1/6$ and the cost-benefit ratio is $c/R=0.005$. The cost-benefit ratio captures the trade-off between the waiting cost of vehicles and their monetized benefits from platooning. The figure shows that vehicles are released in platoons of $6$ or $7$ vehicles. This is consistent with the fact that the optimal release rule, for $\lambda=1/6$ and $c/R=0.005$, is to release the vehicles when the number of vehicles at the hub is more or equal to the threshold $n^*=6$. In the next section, the threshold $n^*$ is computed when the arrival rate $\lambda$ and the cost-benefit ratio $c/R$ are varied.

\subsection{Computing the threshold  $n^*$}

\begin{figure}	
	\centering
	\includegraphics[scale=0.64]{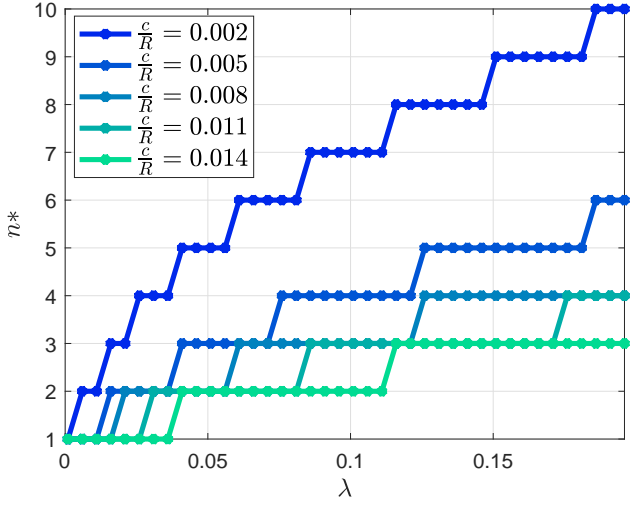}
	\caption{The threshold $n^*$ as a function of $\lambda$ for different values of the cost-benefit ratio $c/R$.}
	\label{SR}
\end{figure}

The optimal release rule in Theorem \ref{thm1} is to release the vehicles in a platoon when the number of vehicles exceeds the threshold $n^*$. Figure \ref{SR} shows the threshold $n^*$ as a function of the arrival rate $\lambda$, for different values of the cost-benefit ratio $c/R$. Figure \ref{SR} shows that the threshold increases when $\lambda$ increases. This is intuitive since the incentive to wait for more vehicles to arrive is higher when more vehicle are expected to arrive. Moreover, the figure also shows that the threshold increases when $c/R$ decreases. This is intuitive since the incentive to stay at the hub is high when the cost of waiting is low and the platooning reward is high.

\subsection{Studying the utility, platoon length and waiting times}

The results presented in this subsection are obtained by $1000$ Monte-Carlo samples. For each sample, the arrivals $x_1,...,x_J$ and the initial number of vehicles $n_0>0$ are re-drawn. The cost-benefit ratio is assumed to be $c/R=0.005$. We compare the performance of the optimal release time rule in Theorem \ref{thm1} against a periodic policy, a spontaneous platooning policy and a policy with the non-causal knowledge of the arrival process. Under the periodic policy, the vehicles that arrive to the hub within each $5  \text{-minute}$ ($60$~time-step) interval are grouped into a platoon. In the non-causal policy, the platoon coordinator has the non-causal knowledge of future arrivals and releases at the time-step which maximizes its reward. In the spontaneous policy, the vehicles are released at the same time-step as they arrive and vehicles platoon spontaneously.

\begin{figure}
	\centering
	\includegraphics[scale=0.64]{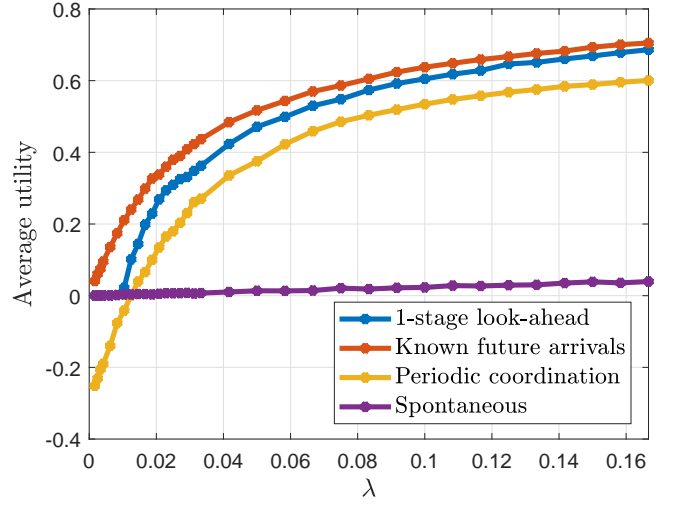}
	\caption{Average utility of the vehicles as a function of $\lambda$.}
	\label{UT}
\end{figure}

Figure \ref{UT} shows the average utility per vehicle, for different release time policies, as a function of the arrival rate $\lambda$ which is varied from $0$ to $1/6$. For each releasing policy, the average utility increases when $\lambda$ increases. According to Figure \ref{UT}, the highest average utility per vehicle is obtained when the coordinator has the knowledge of future arrivals to the hub. This is expected due to the fact that the coordinator has the non-causal knowledge of the future arrivals and always picks the departure time that corresponds to the highest reward. Note that the availability of future arrivals may not always be possible but the non-causal policy provides an upper bound on the performance of the optimal one time-step look-ahead policy. Figure \ref{UT} shows that the performance of the optimal one time-step look-ahead rule is close to that of the non-causal release rule. It also shows that the optimal one time-step look-ahead rule outperforms the spontaneous platooning  and the periodic release policies. Based on~Figure \ref{UT}, the spontaneous platooning policy results in a very low average utility. This suggests that coordination is needed in order to obtain substantial benefit from platooning.  Another observation is that when $\lambda$ is low, the average utility of the periodic release policy is negative. This is due to the fact that the cost of waiting overtakes the average platooning benefit when the arrival rate is low.

\begin{figure}
	\centering
	\includegraphics[scale=0.64]{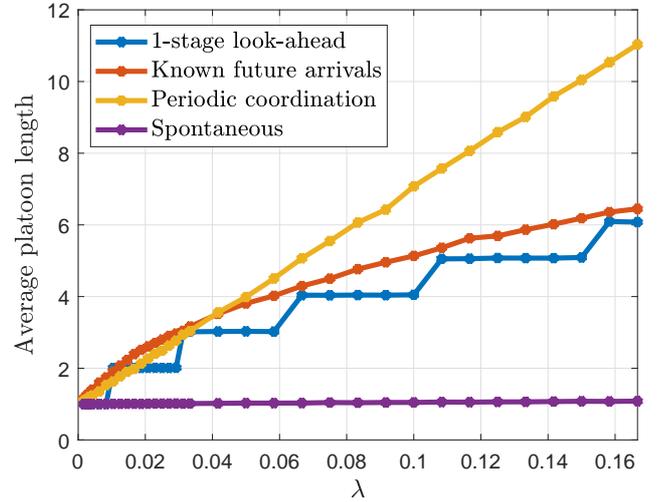}
	\caption{Average platoon length as a function of $\lambda$.}
	\label{Len}
\end{figure}

Figure \ref{Len} shows the average platoon length, for different release policies, as a function of $\lambda$. The vehicles that depart alone are counted as one-vehicle platoons. The figure shows that the average platoon length increases when $\lambda$ increases, for all release policies. This is expected since, in average, more vehicles arrive to the hub at each time-step. Figure \ref{Len} shows that when the optimal release rule is employed, the trajectory of the platoon length is step-shaped. A jump occurs every time $\lambda$ reaches a point where the threshold $n^*$ is increased by one.

\begin{figure}
	\centering
	\includegraphics[scale=0.64]{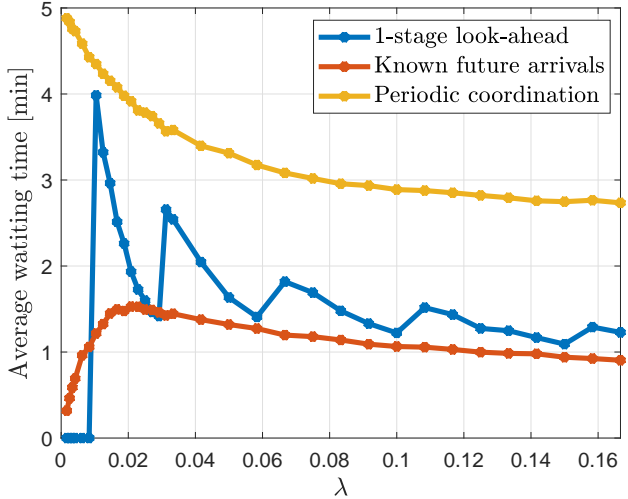}
	\caption{Average waiting time per vehicle as a function of $\lambda$.}
	\label{Wait}
\end{figure}

Figure \ref{Wait} shows the average waiting time per vehicle for different release policies as a function of $\lambda$. The average waiting time of the optimal release rule is saw-tooth shaped. This is because at each point that $\lambda$ reaches a value where the threshold $n^*$ increases by one, the average waiting time jumps to a new value. In the regions in-between the jumps, the average waiting decreases, since when more vehicles arrive to the hub, it takes shorter time to reach the threshold. Moreover, it is worth pointing out that when $\lambda$ is low, the optimal release policy has zero waiting time, similar to the spontaneous policy.

\section{Conclusions}\label{CON}

We have considered the platoon coordination problem where vehicle form platoons at a hub to which vehicles arrive according to a stochastic arrival process.  The coordinator has the knowledge of the statistical distribution of the vehicle arrival process and decides at each time-step if the vehicles at the hub will be released in a platoon. We model the release time problem as a stopping rule problem, where stopping corresponds to releasing vehicles from the hub.

 Under an i.i.d assumption of the vehicle arrival process, we showed that it is optimal for the coordinator to only look one time-step ahead when deciding whether to release the vehicles from the hub or not.
  This was shown by first proving that the release time problem is monotone. We showed that the optimal release rule is in the form of a threshold-based policy in which the coordinator decides to release the vehicles if the number of vehicles located at the hub exceeds a certain threshold. The threshold depends on  the cost of waiting, the benefits from platooning and the statistical distribution of the vehicle arrival process.

The performance of the optimal release time rule was evaluated in a simulation of a potential hub in Sweden. Historical traffic data was used to obtain realistic values of the arrival rates to the hub. In the simulation, it was shown that the optimal release rule performed nearly as good as a non-causal policy where the coordinator has full information about future arriving vehicles. Moreover, the simulation  showed that the optimal release rule outperformed a periodic release rule.

\appendix
\section{} \label{AA}
\begin{pf}[Theorem \ref{thm1}]
	We show that the condition in Definition \ref{mono} holds for the considered release time problem. The one time-step look-ahead release rule calls for releasing at time-step $k$ if 
	\begin{align}\label{sla1a}
	\frac{n_k-1}{n_k}R-ck  \geq   \sum \limits_{x} \frac{n_k+x-1}{n_k+x}R \text{P}(x) -c(k+1).
	\end{align}
	Since $\sum \limits_{x}\text{P}(x)=1$, we have that \eqref{sla1a} is equivalent to
	\begin{align*}
	&\frac{c}{R} \geq   \sum \limits_{x}( \frac{n_k+x-1}{n_k+x}-\frac{n_k-1}{n_k})\text{P}(x),
	\end{align*}
	which can be written as
	\begin{align}\label{sla1b}
	\frac{c}{R} \geq \sum \limits_{x} \frac{x}{n_k^2+n_kx}\text{P}(x).
	\end{align}
	Similarly, it can be shown that, the one time-step look-ahead release rule calls for releasing at time-step $k+1$ if
	\begin{align}\label{sla2b}
	\frac{c}{R} \geq \sum \limits_{x} \frac{x}{n_{k+1}^2+n_{k+1}x}\text{P}(x).
	\end{align}
	 Note that $n_{k}$ and $n_{k+1}$ appear in the right-hand-side of the inequalities \eqref{sla1b} and \eqref{sla2b}, respectively, and this is the only difference between these inequalities. Moreover, the right-hand-side of \eqref{sla1b} and \eqref{sla2b} are decreasing in $n_{k}$ and $n_{k+1}$, respectively. Therefore, if inequality \eqref{sla1b} is satisfied it implies that  inequality \eqref{sla2b} is satisfied if $n_{k+1}\geq n_{k}$. The release time problem is therefore monotone by the fact that $n_{k+1}=n_{k}+x_k\geq n_{k}$.

	\cite{Ferguson2007} showed that it is optimal to look one time-step ahead in finite horizon monotone stopping rule problems. By this fact and by that the considered release time problem is monotone, it follows that the one time-step look-ahead release rule is optimal. Moreover, the one time-step look-ahead release rule calls for releasing at time-step $k$ if the inequality \eqref{sla1b} is satisfied, which it is if and only if $n_k\geq n^*$. The same policy is optimal in the case of infinite horizon, as \cite{Chow1963} showed the optimality of the one time-step look-ahead rule of infinite horizon monotone stopping problems.

 \end{pf}

\bibliography{RefDatabase} 

\begin{thebibliography}{16}
\providecommand{\natexlab}[1]{#1}
\providecommand{\url}[1]{\texttt{#1}}
\providecommand{\urlprefix}{URL }
\expandafter\ifx\csname urlstyle\endcsname\relax
  \providecommand{\doi}[1]{doi:\discretionary{}{}{}#1}\else
  \providecommand{\doi}{doi:\discretionary{}{}{}\begingroup
  \urlstyle{rm}\Url}\fi

\bibitem[{{Adler} et~al.(2016){Adler}, {Miculescu}, and {Karaman}}]{Adler2016}
{Adler}, A., {Miculescu}, D., and {Karaman}, S. (2016).
\newblock Optimal policies for platooning and ride sharing in autonomy-enabled
  transportation.
\newblock In \emph{2016 Workshop on Algorithmic Foundations of Robotics
  (WAFR)}.

\bibitem[{Alam et~al.(2015)Alam, Besselink, Turri, M{\aa}rtensson, and
  Johansson}]{Alam2015}
Alam, A., Besselink, B., Turri, V., M{\aa}rtensson, J., and Johansson, K.H.
  (2015).
\newblock Heavy-duty vehicle platooning for sustainable freight transportation:
  A cooperative method to enhance safety and efficiency.
\newblock \emph{IEEE Control Systems Magazine}, 35(6), 34--56.

\bibitem[{Bhoopalam et~al.(2018)Bhoopalam, Agatz, and Zuidwijk}]{Bhoopalam2018}
Bhoopalam, A.K., Agatz, N., and Zuidwijk, R. (2018).
\newblock Planning of truck platoons: A literature review and directions for
  future research.
\newblock \emph{Transportation Research Part B}, 107, 212--228.

\bibitem[{Boysen et~al.(2018)Boysen, Briskorn, and Schwerdfeger}]{Boysen2018}
Boysen, N., Briskorn, D., and Schwerdfeger, S. (2018).
\newblock The identical-path truck platooning problem.
\newblock \emph{Transportation Research Part B: Methodological}, 109, 26 -- 39.

\bibitem[{Browand et~al.(2004)Browand, McArthur, and Radovich}]{Browand2004}
Browand, F., McArthur, J., and Radovich, C. (2004).
\newblock Fuel saving achieved in the field test of two tandem trucks.
\newblock \emph{Technical report, University of Sourthern California}.

\bibitem[{Chow and Robbins(1963)}]{Chow1963}
Chow, Y.S. and Robbins, H. (1963).
\newblock On optimal stopping rules.
\newblock \emph{Zeitschrift f{\"u}r Wahrscheinlichkeitstheorie und Verwandte
  Gebiete}, 2(1), 33--49.

\bibitem[{Farokhi and Johansson(2013)}]{Farokhi2013}
Farokhi, F. and Johansson, K.H. (2013).
\newblock A game-theoretic framework for studying truck platooning incentives.
\newblock In \emph{16th International IEEE Conference on Intelligent
  Transportation Systems (ITSC 2013)}, 1253--1260.

\bibitem[{{Ferguson}(2007)}]{Ferguson2007}
{Ferguson}, T. (2007).
\newblock Optimal stopping and applications.
\newblock \urlprefix\url{https://www.math.ucla.edu/~tom/Stopping}.

\bibitem[{{Johansson} et~al.(2018){Johansson}, {Nekouei}, {Johansson}, and
  {M{\aa}rtensson}}]{Johansson2018}
{Johansson}, A., {Nekouei}, E., {Johansson}, K.H., and {M{\aa}rtensson}, J.
  (2018).
\newblock Multi-fleet platoon matching: A game-theoretic approach.
\newblock In \emph{2018 21st International Conference on Intelligent
  Transportation Systems (ITSC)}, 2980--2985.

\bibitem[{Larsen et~al.(2019)Larsen, Rich, and Rasmussen}]{Larsen2019}
Larsen, R., Rich, J., and Rasmussen, T.K. (2019).
\newblock Hub-based truck platooning: Potentials and profitability.
\newblock \emph{Transportation Research Part E: Logistics and Transportation
  Review}, 127, 249 -- 264.

\bibitem[{Larsson et~al.(2015)Larsson, Sennton, and Larson}]{Larsson2015}
Larsson, E., Sennton, G., and Larson, J. (2015).
\newblock The vehicle platooning problem: Computational complexity and
  heuristics.
\newblock \emph{Transportation Research Part C: Emerging Technologies}, 60, 258
  -- 277.

\bibitem[{Liang et~al.(2016)Liang, {M{\aa}rtensson}, and Johansson}]{Liang2016}
Liang, K., {M{\aa}rtensson}, J., and Johansson, K.H. (2016).
\newblock Heavy-duty vehicle platoon formation for fuel efficiency.
\newblock \emph{IEEE Transactions on Intelligent Transportation Systems},
  17(4), 1051--1061.

\bibitem[{Tsugawa et~al.(2016)Tsugawa, Jeschke, and Shladover}]{Tsugawa2016}
Tsugawa, S., Jeschke, S., and Shladover, S.E. (2016).
\newblock A review of truck platooning projects for energy savings.
\newblock \emph{IEEE Transactions on Intelligent Vehicles}, 1(1), 68--77.

\bibitem[{van~de Hoef et~al.(2018)van~de Hoef, Johansson, and
  Dimarogonas}]{Hoef2018}
van~de Hoef, S., Johansson, K.H., and Dimarogonas, D.V. (2018).
\newblock Fuel-efficient en route formation of truck platoons.
\newblock \emph{IEEE Transactions on Intelligent Transportation Systems},
  19(1), 102--112.

\bibitem[{Xiong et~al.(2020)Xiong, Sha, and Jin}]{Xiong2020}
Xiong, X., Sha, J., and Jin, L. (2020).
\newblock Optimizing coordinated vehicle platooning: An analytical approach
  based on stochastic dynamic programming.
\newblock \emph{ArXiv}, abs/2003.13067.

\bibitem[{Zhang et~al.(2017)Zhang, Jenelius, and Ma}]{Zhang2017}
Zhang, W., Jenelius, E., and Ma, X. (2017).
\newblock Freight transport platoon coordination and departure time scheduling
  under travel time uncertainty.
\newblock \emph{Transportation Research Part E: Logistics and Transportation
  Review}, 98, 1 -- 23.

\end{thebibliography}
\end{document}